\documentclass[sigconf]{acmart}

\usepackage{soul}
\usepackage{multirow}
\usepackage{caption}
\usepackage{color,soul}
\usepackage{graphicx}

\usepackage{balance}

\begin{document}

% \title{Energy profiling of private 5G}
% \title{Energy profiling from standby-cost to private 5G Functioning in Edge-cloud continuum}
\title{Energy Profiling and Analysis of 5G Private Networks: Evaluating Energy Consumption Patterns}

\author{Johirul Islam}
\email{johirul.islam@oulu.fi}
\orcid{0000-0002-7523-0666}
\affiliation{%
  % \institution{University of Oulu, Finland}
  %\city{Oulu}
  \country{University of Oulu, Finland}
}

\author{Ijaz Ahmad}
\email{ahmad.ijaz@oulu.fi}
\orcid{0000-0002-6152-8947}
\affiliation{%
  % \institution{University of Oulu, Finland}
  %\city{Oulu}
  \country{University of Oulu, Finland}
}

\author{Shakthi Gimhana}
\email{shakthi.hinguralaarachchigedon@oulu.fi}
\orcid{0000-0001-7263-9469}
\affiliation{%
 % \institution{University of Oulu, Finland}
 % \city{Oulu}
 \country{University of Oulu, Finland}
}

\author{Juho Markkula}
\email{juho.markkula@oulu.fi}
\orcid{0000-0001-7402-3414}
\affiliation{%
  % \institution{University of Oulu, Finland}
  %\city{Oulu}
  \country{University of Oulu, Finland}
}

\author{Erkki Harjula}
\email{erkki.harjula@oulu.fi}
\orcid{0000-0001-5331-209X}
\affiliation{%
  % \institution{University of Oulu, Finland}
  %\city{Oulu}
  % \state{Beijing Shi}
  \country{University of Oulu, Finland}
}



\begin{abstract}
%\textcolor{cyan}{[Erkki] You can build the abstract around the following motivation: Private 5G gains attention since, beyond the generic benefits of 5G (low latency, high throughput, slicing, support for IoT, ...) they are owned and operated by the organization itself, providing greater control over security and privacy => Interesting for healthcare. Sustainability and cost considerations emphasize the energy aspect. Hower, practcal knowledge is missing on energy consumption characteristics and the factors affecting it. This paper aims to shed light on this, Blaa...} 

%\textcolor{blue}{Shakthi:}
Private 5G networks provide enhanced security, a wide range of optimized services through network slicing, reduced latency, and support for many IoT devices in a specific area, all under the owner's full control.  Higher security and privacy to protect sensitive data is the most significant advantage of private networks, in e.g., smart hospitals. For long-term sustainability and cost-effectiveness of private 5G networks, analyzing and understanding the energy consumption variation holds a greater significance in reaching toward green private network architecture for 6G. This paper addresses this research gap by providing energy profiling of network components using an experimental laboratory setup that mimics real private 5G networks under various network conditions, which is a missing aspect in the existing literature.
\end{abstract}







\maketitle

\section{Introduction}
%\textcolor{cyan}{[Erkki:] Introduce Private (Standalone) 5G and some examples in industry and academia. Tell what it enables and why it's an interesting development on our way towards 6G. Introduce the requirement for sustainability and tell about the growing energy prices that set requirements to the energy-efficiency of networks and private 5G}

%\textcolor{blue}{Ijaz:} 
Private Standalone 5G networks %also known as standalone 5G, 
are getting more popular as they offer unique advantages such as greater control, enhanced security, and reduced latency. These networks are largely deployed and controlled by organizations, such as healthcare and Industry 4.0 initiatives. In academia, private 5G networks are popular, considering their unique features of ultra-low latency, unparalleled high speed, and are largely deployed as sandbox for experiments pertaining to new communication protocols, emerging authentication and encryption techniques, and thus paving their way towards 6G. A few examples of 5G in smart industries are BMW using 5G for deploying blockchain pilot and Bosch \footnote{https://www.bosch.com/stories/5g-industry-4-0/} deploying private 5G in manufacturing plants to improve automation and increase productivity.
The University of Oulu is leading the EU-sponsored project of Future Hospital - Hola 5G \footnote{https://ouluhealth.fi/hola-5g-oulu-brings-to-healthcare-europes-first-ever-private-5g-sa-network/} aiming to test the applicability, feasibility and efficiency of the private 5G network in The Oulu University hospital. The project equips the hospital with private 5G test network and see its feasibility in future healthcare scenarios such as those involving extended reality (XR) requiring ultra-low latency for real-time sugeries and surgical simulations.\\
Sustainability has become a key aspect of ICT, as there is a global shift toward greener computation and reduced carbon footprint in response to increasing environmental concerns and regulatory requirements. Rising energy prices are forcing companies to look for energy-efficient and cost-effective solutions. Private 5G networks provide an opportunity for energy optimization due to its granulated controls. Considering the data-driven applications of 5G such as Internet of Things, extensive machine-to-machine (M2M), Edge Computing and integration of Artificial Intelligence makes energy optimization a critical factor for success of private 5G networks.

%For example, Hola 5G project

% \textcolor{cyan}{Faheem: }Edge-cloud continuum computing has become more popular because of the exponential growth of the Internet of things (IoT) devices and their real time applications with growing need for data processing \cite{gkonis2023survey}. Nevertheless, this traditional 2-tier edge-cloud architecture is not well fit for constrained IoT devices requiring ultra-low-latency. To address this, authors in \cite{harjula2019decentralized}, considered the local edge computing resources to the traditional 2-tier edge-cloud computing architecture and proposed a 3-tier edge-cloud architecture where computing resources are utilized for different purposes e.g., the local edge (Tier-1) for immediate, low-latency processing; the intermediate edge (Tier-2) for more complex tasks requiring moderate resources; and the cloud (Tier-3) for high-capacity, centralized processing and data storage. However, the overall energy consumption is increasing day-by-day and hence requiring an energy-aware service orchestration component for edge-cloud service architecture. This deliverable provides a comprehensive energy orchestration concept for 3-tier edge-cloud service architecture by considering a system model with a few high-level requirements and a framework for analyzing the service-wise energy-consumption to build and develop an energy-aware service orchestration platform.
\section{Background \& Related Work}
%\textcolor{cyan}{[Erkki:]This section should focus on showing that there is a gap in the literature (is there?) regarding the granular  energy profiling in private 5g networks}
%\textcolor{blue}{Shakthi:}
With the energy harvesting methods gaining attention and emerging research on wireless power transfer to network elements, energy profiling of the network components under different traffic patterns requires a more in-depth research focus \cite{mughees2021energy}. Moreover, the Open radio access network (O-RAN) emerging as the entry point for 6G, addressing the security of the open interfaces of O-RAN is leading to increased energy consumption \cite{porambage2023xcaret}. Element-wise energy profiling under varying conditions is key for designing green security policies for O-RAN networks.
% \begin{figure}[!h]
%   \centering
%   % trim => left, down, right and top
%   \includegraphics[trim=0.29in 0.30in 0.23in 0.26in, clip, width=\linewidth]{figures/experiment-setup-3.pdf}
%   \caption{Experiment Setup.}
%   \label{fig:experiment_setup}
%   \Description{Energy measurement framework.}
% \end{figure}
%\textcolor{blue}{Ijaz:} 
\noindent
%The study on Massive MIMO systems \cite{khwandah2021massive} highlights the importance of beamforming and energy efficiency in public 5G networks, but lacks specific insights into private 5G setups and their unique configurations in real-world deployments.
The research on energy efficiency in adaptive massive MIMO networks \cite{salah2022energy} \cite{khwandah2021massive} discuss overall energy savings and optimization strategies, but do not highlight the breakdown of energy consumption in individual network components such as core, radio and edge separately. The existing studies on adaptive massive MIMO systems, explore energy optimization methods using genetic algorithms, but mainly in the context of public 5G networks. There is an opportunity to adapt and test these techniques specifically for private 5G networks, which often have different traffic patterns and requirements than public networks.
% \textcolor{blue}{Shakthi:} 
This research focuses on addressing the research gap of energy profiling under varying network conditions with an experimental network setup to encourage the research community to conduct more network-level energy profiling research which can be highly beneficial in designing energy-efficient 6G network architecture.

\section{Methodology}
%\textcolor{cyan}{[Erkki:] Introduce the test setup. Oaibox, usrp, Netio measurement HW\&SW. Very briefly. Photo if there is space.}

\begin{figure}[!h]
  % \vspace{-5mm}
  \centering
  % trim => left, down, right and top
  \includegraphics[trim=0.29in 0.30in 0.23in 0.26in, clip, width=\linewidth]{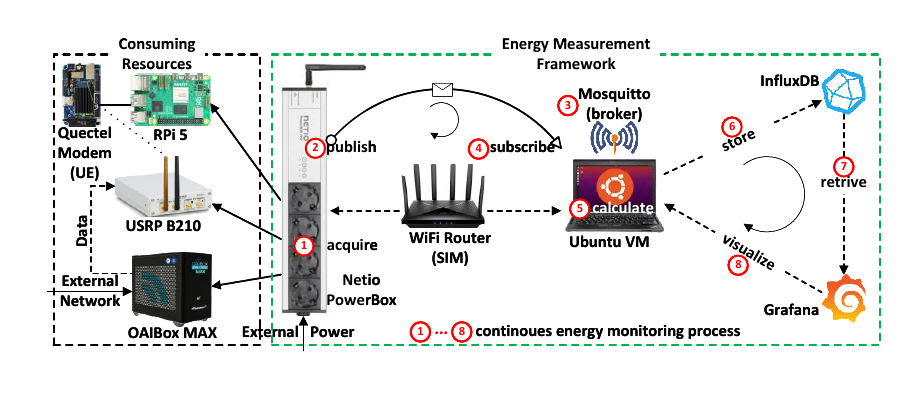}
  \caption{Experiment Setup.}
  \label{fig:experiment_setup}
  \Description{Energy measurement framework.}
  % \vspace{-5mm}
\end{figure}

% The energy profiling measurements are done with Netio.
%\textcolor{red}{Johirul:} 
The energy profiling measurements are done with Netio PowerBox 4KF. In the experiment, the MQTT protocol is used to obtain the associated data from Netio. The general data collection, storage, and visualization process is shown in Figure~\ref{fig:experiment_setup} with various steps.
% It has an ethernet port and a WiFi interface and can connect to the internet and share data with other IoT nodes. However, the energy consumption of an electric component can be measured through a specific socket of the Netio device through which the electrical component is getting the power from Netio to run it. However, the power utilization of that electric component can be acquired from the Netio either periodically or based on some events e.g., voltage/current level changes in a certain socket. 
% At first, the voltage (V), current (I), and true power factor (pf) are obtained from the Netio for a target device that is connected to the Netio's certain socket. 
% Then the power consumption of that target device is calculated manually with the equation~\ref{eqn:power}. 
% MQTT PAYLOAD: \\ 
% JSON ${\Rightarrow} \{ "current": I, "voltage": V, "power\_factor": pf\}$
% EQUATION: \\
% P = V * I \\
% E = pf * P * ${\Delta t}$ \\
% \begin{equation}
% \label{eqn:power}
% P = V * I * pf
% \end{equation}
% \begin{equation}
% \label{eqn:energy}
% E =  P * {\Delta t}
% \end{equation}

% To share the measured data with the other IoT nodes, Netio supports various open APIs e.g., SNMP, Modbus/TCP, MQTT, Telnet, etc. 
% In the experiment, the MQTT protocol is used to obtain the associated data from Netio. The general data collection, storage, and visualization process is shown in Figure~\ref{fig:experiment_setup} with various steps. 
\noindent
At first, Netio acquires the the voltage (V), current (I), and true power factor (pf) for a target device that is connected to the Netio's certain socket (step 1) and then publishes (step 2) these data to a Ubuntu VM (step 3-4), which runs in a laptop. Later, the Ubuntu VM extracts and manually calculates power with the equation~\ref{eqn:power} (step 5) and stores these in InfluxDB (step 6).
\begin{equation}
\label{eqn:power}
P = V * I * pf
\end{equation}
Finally, Grafana retrieves the data from InfluxDB (step 7) and visualizes the data through a browser (step 8). All these 1-8 are a continuous process of our energy measurement framework.

In the lab, OAIBox Max and Universal Software Radio Peripheral (USRP) B210 are used to analyze the energy consumption of the private 5G networks whereas the OAIBox Max includes the core network 5G (CN5G) and gNB (gNode B) while the USRP B210 acts as a radio unit of gNB. The overall hardware and software components that are used in the lab are discussed below.

\textbf{Hardware components:} (1) \textit{OAIBox Max} is running with Ubuntu 22.04 having a 5.7 GHz AMD Ryzen CPU 
% (24 cores) 
with 32 GB RAM and 1 TB storage. (2) \textit{USRP B210} is a dual-channel transceiver that covers continuous frequency from 70 MHz – 6 GHz. (3) \textit{Raspberry Pi 5 (RPi 5)} is running with Debian GNU/Linux 12 having a 2.4 GHz ARM Cortex A76  CPU 
% (4 cores) 
with 8 GB RAM and 32 GB storage. (4) \textit{Quectel Modem (RM500-GL)} with 5G enabled SIM (from Allbesmart) that acts as a UE that is getting radio signal from USRP B210. (5) \textit{Ubuntu 24.04 VM} has an Intel Core i7 2.40 GHz CPU 
% (8 v-cores) 
with 8 GB RAM and 25 GB storage. (6) \textit{Netio PowerBox 4KF} that has 4 sockets and includes both LAN and Wi-Fi.

\textbf{Software components:} (1) Both the \textit{mosquitto} and \textit{mosquitto-clients} of version \textit{2.0.18} are installed for MQTT broker and client at the Ubuntu VM (2) \textit{InfluxDB 2.7.10} server and CLI are installed in Ubuntu VM for storing the Netio measured data (3) \textit{Grafana 11.2.2} is installed in Ubuntu VM as a visualizer to present the energy data in graph format (4) \textit{iPerf 3.16} is installed in both the OAIBox Max and RPi 5 to evaluate the performance of the private 5G network whereas OAIBox Max is considered as a client and the RPi 5 is considered as a server.

\section{Energy profiles}
%\textcolor{cyan}{[Erkki:] Energy profiles (consumption figures for core, radio (ue connected/disconnected), uplink\&downlink\&idle}

\begin{figure}[!h]
  % \vspace{-1mm}
  \centering
  \includegraphics[trim=0in 0.25in 0in 0in, clip, width=\linewidth, height=3.3cm]{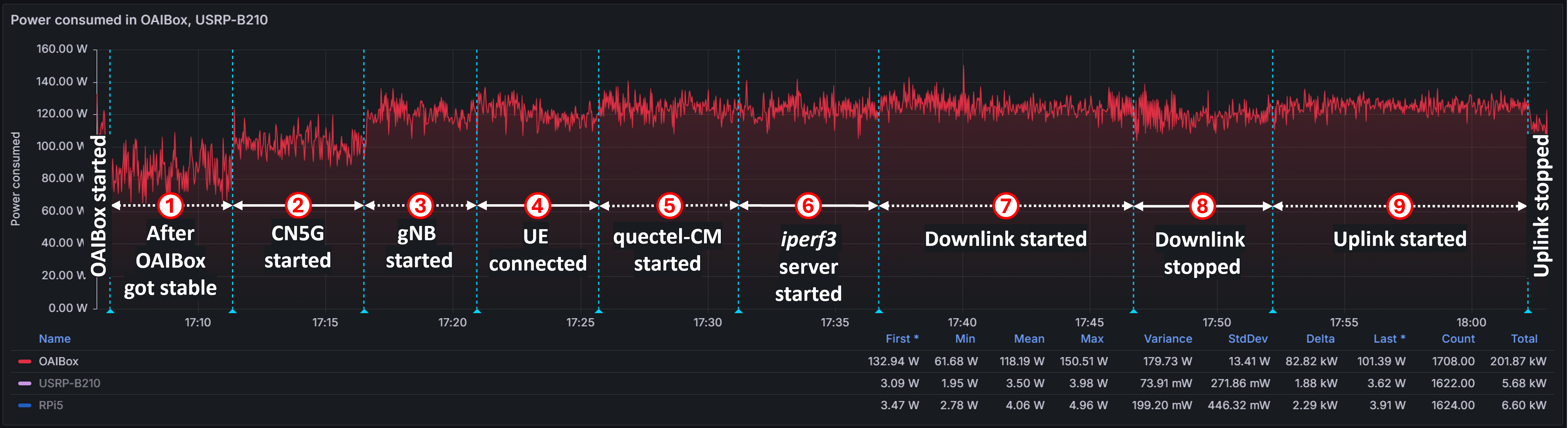}
  \caption{Energy consumption at 20 MHz.}
  \label{fig:energy-at-20MHz}
  \Description{Energy consumption at 40 MHz.}
  % \vspace{-1mm}
\end{figure}

% To measure the energy consumption of the private 5G network, 
We observe the energy consumption of OAIBox (5G core) and USRP B210 (5G radio) at different stages for the 20 MHz 
% (with 3 downlink, 1 free, and 1 uplink TDD slots)
% such as when the - (1) OAIBox got stable, (2) CN5G started, (3) gNB started, (4) UE connected, (5) modem started, (6) iperf3 started, (7) downlink, (8) downlink stopped, and (8) uplink 
as shown in Figure~\ref{fig:energy-at-20MHz}. It is observed that in the idle mode (step 1), OAIBox and USRP consume roughly 6.91 Wh and 0.23 Wh respectively. 
% while they are idle i.e., no other 5G component e.g., 5G core, gNB, etc are not running.
For CN5G (step 2), the energy increases by roughly 1.93 Wh in the OAIBox once it starts. 
Additionally, 0.04 Wh and 0.03 Wh are increased both in OAIBox and USRP respectively for gNB while it starts (step 3). 
In the current setup, Quectel UE  gets power from the OAIBox. It increases by 0.81 Wh and 0.02 Wh at the OAIBox and USRP respectively while it is connected or plugged (step 4) but not active for exchanging data.

\noindent
The \textit{quectel-CM} activates the modem for exchanging data through 5G SIM. At the active state (step 5), it rises 1.68 Wh and 0.04 Wh both at OAIBox and USRP respectively.
The \textit{iperf3} server which started on RPi 5 (step 6) does not have any effect on energy cost neither on the OAIBox and USRP. 
However, during the downlink (step 7), \textit{iperf3} client increases 9.58 Wh and 0.30 Wh at the OAIBox and USRP and returns (step 8) to the previous state as observed in step 6. On the other hand, \textit{iperf3} client again increases 9.63 Wh and 0.28 Wh at the OAIBox and USRP during the uplink (step 9).

%\subsection{Energy-aware}
%\textcolor{cyan}{Start text . . . .}

%\section{Evaluation and Results}
%\textcolor{cyan}{Start text . . . .}

\section{Discussion and conclusions}
%\textcolor{cyan}{[Erkki:] We have in this paper provided energy profiles for the private 5G core+radio lab setup. The results give other researchers and industry a glance to the factors affecting the energy consumption in practical private 5G setups. \textbf{Limitations:} these results concern a controlled lab setup, with 1 access point and 1(-2?) ue:s. Real-world scenarios are more complex. This also concerns only the networking side, not computational side (edge etc.). \textbf{Future work:} More realistic scenario to be implemented in the Hola 5G in near future, still experimental network, but in real hospital environment. Future work also includes profiling the energy on UE side and the computational energy (ue\&edge nodes).}

%\textcolor{blue}{Shakthi:}
We have provided an experimental analysis of the energy profiling for the 5G core %OAIBox (CN5G+gNB) 
and 5G radio units %USRP B210 
under varying network conditions that provide a look at the factors that affect energy consumption in a real private 5G network. The energy consumption of the radio unit caused by the traffic can be directly mapped with the traffic that will be input to the open fronthaul of the O-RAN. One of the main limitations of our work is that real-world private 5G network conditions can be more complex and have more access points compared to our laboratory setup with only one access point.  Our future work will include conducting energy profiling on a real-world private 5G network with many access points in a real hospital under the Hola 5G project
% \footnote{\textbf{Acknowledgments:} This work is supported by Business Finland via Eware-6G project (grant 8819/31/2022), the Research Council of Finland through 6G Flagship (grant 318927), and Connecting Europe Facility (CEF) funded Hola 5G project. 
% % We express our gratitude to Juho Markkula, eware-6g project manager, for his valuable assistance in conducting the experiments.
% } 
also considering the energy consumption of UEs and edge nodes.

% \begin{table}
%   \caption{Frequency of Special Characters}
%   \label{tab:freq}
%   \begin{tabular}{ccl}
%     \toprule
%     Non-English or Math&Frequency&Comments\\
%     \midrule
%     \O & 1 in 1,000& For Swedish names\\
%     $\pi$ & 1 in 5& Common in math\\
%     \$ & 4 in 5 & Used in business\\
%     $\Psi^2_1$ & 1 in 40,000& Unexplained usage\\
%   \bottomrule
% \end{tabular}
% \end{table}

% \begin{table*}
%   \caption{Some Typical Commands}
%   \label{tab:commands}
%   \begin{tabular}{ccl}
%     \toprule
%     Command &A Number & Comments\\
%     \midrule
%     \texttt{{\char'134}author} & 100& Author \\
%     \texttt{{\char'134}table}& 300 & For tables\\
%     \texttt{{\char'134}table*}& 400& For wider tables\\
%     \bottomrule
%   \end{tabular}
% \end{table*}

% \section{Figures}
% \begin{figure}[h]
%   \centering
%   \includegraphics[width=\linewidth]{sample-franklin}
%   \caption{1907 Franklin Model D roadster. Photograph by Harris \&
%     Ewing, Inc. [Public domain], via Wikimedia
%     Commons. (\url{https://goo.gl/VLCRBB}).}
%   \Description{A woman and a girl in white dresses sit in an open car.}
% \end{figure}

\section{Acknowledgments}
This work is supported by Business Finland via Eware-6G project (grant 8819/31/2022), the Research Council of Finland through 6G Flagship (grant 318927), and Connecting Europe Facility (CEF) funded Hola 5G project.

\balance

\bibliographystyle{ACM-Reference-Format}
\bibliography{references}


\begin{thebibliography}{4}

%%% the Web address.

\ifx \showCODEN    \undefined \def \showCODEN     #1{\unskip}     \fi
\ifx \showDOI      \undefined \def \showDOI       #1{#1}\fi
\ifx \showISBNx    \undefined \def \showISBNx     #1{\unskip}     \fi
\ifx \showISBNxiii \undefined \def \showISBNxiii  #1{\unskip}     \fi
\ifx \showISSN     \undefined \def \showISSN      #1{\unskip}     \fi
\ifx \showLCCN     \undefined \def \showLCCN      #1{\unskip}     \fi
\ifx \shownote     \undefined \def \shownote      #1{#1}          \fi
\ifx \showarticletitle \undefined \def \showarticletitle #1{#1}   \fi
\ifx \showURL      \undefined \def \showURL       {\relax}        \fi
% invisible to TeX
\providecommand\bibfield[2]{#2}
\providecommand\bibinfo[2]{#2}
\providecommand\natexlab[1]{#1}
\providecommand\showeprint[2][]{arXiv:#2}

\bibitem[Khwandah et~al\mbox{.}(2021)]%
        {khwandah2021massive}
\bibfield{author}{\bibinfo{person}{Sinan~A Khwandah}, \bibinfo{person}{John~P Cosmas}, \bibinfo{person}{Pavlos~I Lazaridis}, \bibinfo{person}{Zaharias~D Zaharis}, {and} \bibinfo{person}{Ioannis~P Chochliouros}.} \bibinfo{year}{2021}\natexlab{}.
\newblock \showarticletitle{Massive MIMO systems for 5G communications}.
\newblock \bibinfo{journal}{\emph{Wireless Personal Communications}} \bibinfo{volume}{120}, \bibinfo{number}{3} (\bibinfo{year}{2021}), \bibinfo{pages}{2101--2115}.
\newblock


\bibitem[Mughees et~al\mbox{.}(2021)]%
        {mughees2021energy}
\bibfield{author}{\bibinfo{person}{Amna Mughees}, \bibinfo{person}{Mohammad Tahir}, \bibinfo{person}{Muhammad~Aman Sheikh}, {and} \bibinfo{person}{Abdul Ahad}.} \bibinfo{year}{2021}\natexlab{}.
\newblock \showarticletitle{Energy-efficient ultra-dense 5G networks: recent advances, taxonomy and future research directions}.
\newblock \bibinfo{journal}{\emph{IEEE Access}}  \bibinfo{volume}{9} (\bibinfo{year}{2021}), \bibinfo{pages}{147692--147716}.
\newblock


\bibitem[Porambage et~al\mbox{.}(2023)]%
        {porambage2023xcaret}
\bibfield{author}{\bibinfo{person}{Pawani Porambage}, \bibinfo{person}{Jarno Pinola}, \bibinfo{person}{Yasintha Rumesh}, \bibinfo{person}{Chen Tao}, {and} \bibinfo{person}{Jyrki Huusko}.} \bibinfo{year}{2023}\natexlab{}.
\newblock \showarticletitle{Xcaret: Xai based green security architecture for resilient open radio access networks in 6g}. In \bibinfo{booktitle}{\emph{2023 Joint European Conference on Networks and Communications \& 6G Summit (EuCNC/6G Summit)}}. IEEE, \bibinfo{pages}{699--704}.
\newblock


\bibitem[Salah et~al\mbox{.}(2022)]%
        {salah2022energy}
\bibfield{author}{\bibinfo{person}{Ibrahim Salah}, \bibinfo{person}{M~Mourad Mabrook}, \bibinfo{person}{Kamel~Hussein Rahouma}, {and} \bibinfo{person}{Aziza~I Hussein}.} \bibinfo{year}{2022}\natexlab{}.
\newblock \showarticletitle{Energy efficiency optimization in adaptive massive MIMO networks for 5G applications using genetic algorithm}.
\newblock \bibinfo{journal}{\emph{Optical and Quantum Electronics}} \bibinfo{volume}{54}, \bibinfo{number}{2} (\bibinfo{year}{2022}), \bibinfo{pages}{125}.
\newblock


\end{thebibliography}
\end{document}